\documentclass[%
reprint,amsmath,amssymb,aps,
]{revtex4-1}

\usepackage{graphicx}
\usepackage{dcolumn}
\usepackage{bm}
\usepackage{natbib}
\usepackage{hyperref}
\usepackage{phoenician}
\begin{document}


\title{Pion-Nucleus Elastic Scattering, DCX and Sub-Threshold Resonances}
\author{Mikhail Khankhasayev}
\affiliation{Physics Department,Florida A{\&}M University}
 \email{mikhail.khankhasayev@gmail.com}

\date{\today}

\begin{abstract}
In the scattering of positive pions by nuclei the double-charge-exchange (DCX) reaction creates the possibility of formation of pionic atom states in the vicinity of the threshold of this reaction. These quasi stationary states can manifest themselves as resonances in the elastic scattering cross section. The strength and shape of these resonances is strongly affected by an instability of the nucleus created in the DCX channel. It is shown that this mechanism can explain oscillation structures in the excitation function observed in scattering of low-energy positive pions from $^{12}\text{C}$.      
\end{abstract}

\pacs{25.80.Dj}
\keywords{pion scattering, pion charge-exchange,  mesonic atoms}

\maketitle

 
\section{\label{sec:Sec1}Introduction}

A systematic measurement of energy dependence of the elastic scattering of positive pions from  $^{12}C$ presented in Ref.\cite{Alex98} showed oscillations in the excitation function at energies around 35 MeV. The differential cross sections were measured at six scattering angles $(37^{\circ}, 65^{\circ}, 83^{\circ}, 103^{\circ}, 118^{\circ}, 142^{\circ})$ in the energy range of 18-44 MeV  with an increment in the incident energy of 2 MeV. The differential cross sections were compared to the calculations made within the framework of the unitary scattering theory (UST) of the pion-nucleus scattering developed in Ref. \cite{MKH89}. Despite good description of the measured differential cross sections, the experimental data presented in terms of the excitation function (differential cross section at a given scattering angle as a function of energy) showed oscillation structures which the UST approach did not reproduce. These oscillations become more pronounced at angles around $90^{\circ}$. A typical disagreement between theory and experiment for the excitation function at $83^{\circ}$ is shown in Figure \ref{fig: fig1} by the dashed line. One of the possible explanations of these oscillations could be the formation of a diproton resonance in the $\pi NN$ system. However, a systematic experimental study \cite{Pasyuk97} of the $\pi ^{+}+d \rightarrow p+p$ reaction in the same energy range of 18 - 44 MeV with the increment of 1 - 2 MeV did not show any oscillations in either the total and differential cross sections. 

In the present paper we explore the possibility of explaining these oscillations by the formation of quasi stationary pionic atom states in the vicinity of the threshold of the DCX reaction channel. The DCX $^{12}\text{C}(\pi ^{+},\pi ^{-})^12\text{O}$ reaction creates two opposite charged particles that can form pionic atom states below the threshold of this reaction. If a $\pi^{-}$ forms a bound state with $^{12}\text{O}$ it cannot escape below the threshold due to lack of energy. It is known that such quasi-bound states can manifest themselves as resonances in the elastic cross section. This type of sub-threshold resonances was first investigated and described by A. Baz' \cite{Baz59,Baz71}. Another argument in favor of this mechanism comes from the fact that the mass excess of $^{12}\text{O}$ is 32.06 MeV \cite{Ajzen85}. This brings the Q-value of the DCX channel in $\pi^{+}-^{12}\text{C}$ scattering into the $\sim 30$ MeV energy region. It is worthwhile to note that one of the first experimental observations of $^{12}\text{O}$ was obtained in the DCX reaction $^{12}\text{C}(\pi^{+},\pi^{-})^{12}\text{O}$ \cite{Mord85}.         

The paper is organized as follows. In Section II we present a short overview the UST approach which is used for the description of pion-nucleus interaction. Section III is devoted to the theory of sub-threshold resonances due to formation of quasi-bound pionic atom states. In Section IV, a systematic comparison of theoretical calculations with the data is presented. In Section V we discuss the main results of the paper.              
 
\section{\label{sec:Sec2}Theory}  
In this section we briefly summarize the UST formalism \cite{MKH89} that we use in the description of pion-nucleus scattering. For simplicity we consider the scattering of pions by the nuclei with zero spin. The pion-nucleus scattering amplitude is presented in a standard way as
\begin{equation}
f_{\pi A}= f_{C} +  f_{sc}, 
\label{eq:S1}
\end{equation}
where $f_{C}$ is the Coulomb amplitude, and $f_{sc}$ is the nuclear-Coulomb amplitude        
\begin{equation}
f_{sc}= \frac {1}{2ik}\sum_{l=0}^{\infty}e^{2i\sigma^{\pm}_l}(S_le^{2i\delta^{\pm}_{R,l}} - 1)   
\label{eq:S2}
\end{equation}
where $\sigma^{\pm}_l$ are the coulomb phases and $\delta^{\pm}_{R,l}$ - the Coulomb corrections caused by the effects of the Coulomb distortion of the pion wave. A detailed procedure for calculating these corrections is given in Ref.\cite{MKH89}. The hadronic part is presented by the S- matrix, 
\begin{equation}
S_l = e^{2i\delta _{\pi A,l}},
\label{eq:S3}   
\end{equation}
where $\delta _{\pi A,l}$ are pure hadronic phase shifts which we calculate within the framework of the UST approach \cite{MKH89}.    

The UST approach is  based on the method of evolution with respect to the coupling constant \cite{DAK65,DAK79}. The basic equations are formulated for calculation of the pion-nucleus phase shifts.
\begin{equation}
{\delta}_{\pi A}(k)= {\delta}^{pot}_{\pi A}(k)~+~{\delta}^{abs}_{\pi A}(k).
\label{eq:S4}
\end{equation}
Here, ${\delta}^{pot}$ is the part of the pion-nucleus phase shift that is formed by the multiple scattering of a pion by the nuclear nucleons, and ${\delta}^{abs}$ is the absorption correction. The "potential" part is expressed in terms of the pion-nucleon phase shifts and the nuclear ground state characteristics such as nuclear form factor and correlation functions. The absorption part is expressed in terms of the absorption parameters ${\tilde B}_0$ and ${\tilde C}_0$,
\begin{equation}
{\delta}^{abs}_{\pi A}(k) = A(A-1)\frac{1+\epsilon}{1+2\epsilon/A}
{\hat \rho}^2(\vec q)
[{\tilde B}_0(k)+{\tilde C}_0(k)({\vec{\kappa '}}\cdot {\vec \kappa})],
\label{eq:S5}
\end{equation}
where, $\epsilon={{\omega}_\pi}(k)/2M$, ${\omega}_\pi$ is the pion energy, M is the mass of a nucleon, ${\hat \rho}^2(\vec q)$ is the Fourier transform of the square of nuclear density $\rho (r)$ normalized to unity, ${\vec q}={\vec k}'- {\vec k}$ is the momentum transfer, and ${\vec \kappa}$ and ${\vec \kappa}'$ are the pion momenta in the $\pi-2N$ center-of-mass system. The absorption parameters determined from the pionic atom data are \cite{MKH89},

\begin{eqnarray}
{\tilde B}_0(k)~& =& ~(-0.1~+~i0.1)~\text{fm}^4, \nonumber \\
{\tilde C}_0(k)~& = &~(-2.8~+~i1.0)~\text{fm}^6
\label{eq:S6}
\end{eqnarray}
In its standard form the UST formalism does not take in to account the possibility of formation of sub-threshold resonances in pion-nucleus interactions.    

\section{\label{sec:Sec3}DCX and Sub-threshold Resonances }   
The opening of the DCX reaction channel 
\begin{equation}
\pi^{+} + ^{12}C\rightarrow (\pi^{-}, ^{12}O)^{*}\rightarrow\pi^{+} + ^{12}C , 
\label{eq:S7}
\end{equation}
creates the possibility of formation of bound pionic atom states below the threshold of this reaction. As it was shown by Baz' \cite{Baz59,Baz71}, the formation of such sub-threshold bound states is reflected by creating resonances in the elastic cross section.      
In Refs.\cite{Baz59,Baz71}, a general case of elastic scattering of two particles $X(a,a)X$  below the threshold of the inelastic channel $X(a,b)Y$ when particles $b$ and $Y$ can form bound states was considered. The main idea of theoretical description of the effect of sub-threshold resonances is based on the fact that one can neglect the energy dependence of the wave functions of $(a,X)$ and $(b,Y)$ systems arising from the strong interaction and focus on the analysis of energy dependence of the Coulomb wave function of the bound $(b,Y)$ system.  
   
The formation of the quasi-bound states due to opening of the DCX channel modifies the partial S-matrix (\ref{eq:S3}) in the following way      
\begin {eqnarray}
S_l=e^{2i(\delta_{\pi A, l}+\delta^{res}_l) },
\label{eq:S8} 
\end{eqnarray}    
where the resonance part is given by  
\begin {eqnarray}
\delta ^{res}_l = \arctan\frac{\delta_2 +\kappa_2(-1)^l\zeta _{l} cot\pi \eta}{\delta_1 +\kappa_1(-1)^l\zeta _l  \cot \pi \eta},  
\label{eq:S9} 
\end{eqnarray}    
\begin {eqnarray}
\zeta _{l} = \frac{\pi(2kr  \eta)^{2l+1}}{(2l+1)(\Gamma^2(2l+1)},
\label{eq:S10}    
\end{eqnarray}
where $\eta ={Z\alpha}/{\beta}$ is the Sommerfeld parameter: $Z$ is the nuclear charge, $\alpha = 1/137$, and $\beta$ is the pion velocity in the pion-nucleus c.m. Below the threshold ($E < E_{thr}$) this parameter is given by      
\begin {eqnarray}
\eta=Z\alpha\sqrt{\frac{{\mu}_{\pi A}}{2\vert{E-E_{thr}}\vert}},
\label{eq:S11} 
\end{eqnarray}     
where ${\mu}_{\pi A}$ is the pion-nucleus reduced mass, parameters $\delta_{l,2}$ and $\kappa _{l,2}$ are the real and imaginary parts of energy independent complex parameters  $\delta$ and $\kappa$ in the vicinity of the threshold energy $E_{thr}$. These constants are expressed in terms of the logarithmic derivatives of pion-nucleus wave functions in the strong interaction region. The resonance energies are determined by the following condition: $\delta^{res}=n\pi+\frac{\pi}{2}$, which gives the following equation   
\begin{equation}
\cot \pi X =(-1)^l\frac{\delta_1}{\zeta _l \kappa_1}.
\label{eq:S12}
\end{equation}

The solution of this equation can be written as  
\begin{eqnarray}
E^{res}_{nl} &=& E_{thr}- \frac{Z^2e^4{\mu}_{\pi A}}{2n^2\xi^2_{nl}}, \nonumber \\
\xi _{nl} &\equiv 1& +\frac{1}{n\pi}\arctan{(-1)^l\frac{\zeta_l \beta_1}{\alpha_1}},  n=1,2,3, ... .
\label{eq:S13}
\end{eqnarray}
Here, $\xi _{nl}$ represents the strong interaction shift of pure Coulomb pionic atom energy levels, 
\begin{equation}
E^C_n= -\frac{Z^2e^4{\mu}_{\pi A}}{2n^2}.
\label{eq:S14}       
\end{equation}
From Eq.(\ref{eq:S13}) it follows that for a given total pion-nucleus angular momentum ($l$) there is an infinite number of resonances with increasing density as $E \rightarrow E_{thr}$. It is easy to see that the width of the resonance region is determined by the energy of the first Coulomb level. For reaction (\ref{eq:S7}) the energy of the first pionic atom level is given by     
\begin{equation}
E^C_1=-\frac{{\mu _{\pi N}}(Z+2)^2e^4}{2}.
\label{eq:S15}
\end{equation} 

In the vicinity of each resonance energy $E=E^{res}_{nl}$ the S-matrix can be approximated by the Breit-Wigner formula as   
\begin {equation}  
S_l \approx e^{2i{\delta_{\pi A,l}}}{\bigl(1-\sum_{n=1}^{\infty}\frac{i{\Gamma}^{e}_{nl}}{E-E^{res}_{nl}-i\Gamma^{e}_{nl}/2}\bigr)},  
\label{eq:S16}
\end{equation}     
where ${\Gamma}^e_{nl}$ is the elastic width of each resonance. In applications to real processes the upper limit should be replaced by some finite number N which is determined by experimental energy resolution.   
 
A simple procedure of generalization this formalism to include an important case when one of the particles created in the opening reaction channel is unstable was proposed in Ref. \cite{Baz59}. In this case Eq.(\ref{eq:S16}) is replaced by 
\begin {equation} 
S_l \approx e^{2i{\delta_{\pi A,l}}}{\bigl(1-\sum_{n=1}^{N}\frac{i{\Gamma}^{e}_{nl}}{E-E^{res}_{n,l}-i(\Gamma^{e}_{nl}+\Gamma)/2}\bigr)}, 
\label{eq:S17}
\end{equation}  
where $\Gamma$ represents the particle's energy width. In the considered case this particle is the nucleus created in the DCX reaction.         

Formula (\ref{eq:S16}) can be simplified if the width of created nucleus is much bigger than the elastic widths of corresponding sub-threshold resonances. Indeed, if $\Gamma  >> \Gamma^{e}_{nl}$ one can neglect quantities $\Gamma^{e}_{nl}$ in the denominators and present Eq.(\ref{eq:S16}) in the following form    
\begin {equation} 
S_l \approx e^{2i{\delta_{\pi A,l}}}{\bigl(1-\frac{i\Gamma^e_{tot}}{E-E^{res}_l-i\Gamma/2}\bigr)}, 
\label{eq:S18}
\end{equation} 
where
\begin {equation} 
\Gamma^e_{tot}=\sum_{n=1}^{\infty}\Gamma^{e}_{nl},
\label{eq:S18a}
\end{equation}
and $E^{res}_l$ is some average value of the sub-threshold resonance energy. This formula can be re-written as       
\begin {equation} 
S_l \approx e^{2i{\delta_{\pi A,l}}}{\bigl(1- \gamma\frac{i{\Gamma}}{E-{E^{res}_l}-i(\Gamma /2 }\bigr)},  
\label{eq:S18} 
\end{equation}  
where $\gamma\equiv \Gamma^e_{tot}/\Gamma$. The derived formula presents a single-term Breit-Wigner approximation for the infinite series of the sub-threshold resonances contributing to the scattering process. It is important to note that $\Gamma^e_{tot}$ in an effective elastic width representing the contribution from all resonances at a given orbital momentum $l$.        

\section{\label{sec:Sec3}Calculations}
In scattering of positive pions from $^{12}\text{C}$ the DCX reaction creates two opposite charged particles $(\pi^{-},^{12}O)$ which can form the pionic atom below the threshold of this reaction. This reaction has a positive Q-value ($32.06$ MeV). In addition, a positive pion needs to overcome the nucleus Coulomb barrier. Therefore, the threshold energy of the DCX reaction channel will be determined by the sum of the reaction Q-value and the magnitude of the Coulomb repulsion barrier $\delta V_C$, i.e. $E_{thr}\approx Q+ \delta V_{C}$, where $\delta V_{C}=Ze^2/R, R=r_0 A^{1/3}, r_0=1.1fm$. For $(\pi^{+},^{12}\text{C})$, $\delta V_{C}\approx 3.43$ MeV. Therefore, the threshold energy $E^{thr}\approx 35.45$ MeV. The resonance energies are shifted down from the threshold energy by the amount of the corresponding binding energy in accordance with Eq.(\ref{eq:S13}). The lowest resonance energy corresponds to the $1s$-state in the $(\pi^{-},^{12}\text{O}$ atom. Using Eq.(\ref{eq:S15})one can obtain that $E^C_1\approx -0.23$ MeV, and the following value for the resonance energy $E^{res}_{1s}\approx 35.26$ MeV. In the calculations below this value will be used as the average sub-threshold resonance energy in Eq.(\ref{eq:S18}) as well.          
  
The $^{12}\text{O}$ nucleus is unstable. The width of the g.s. of unbound $^{12}\text{O}$ is known with a large uncertainty: $\Gamma = 0.40 \pm 0.25$ MeV \cite{Ajzen85}. The main decay mode is two-proton emission to the ground state of $^{10}$C. The "elastic" strong interaction width of the $1s$-state of ($\pi^{-},^{12}\text{O}$) is about $10^{-3}$ MeV (see, e.g. \cite{Swanner}). Since the elastic width is much smaller than the width $^{12}\text{O}$ one can use the derived one-term Breit-Wigner approximation (\ref{eq:S18}).          

The spin and parity of $^{12}\text{O}$ g.s. is $0^{+}$. Therefore, the sub-threshold s-resonances can be generated by the pion s-wave only. The s-wave S-matrix is given by     
\begin {equation}
S_0 \approx e^{2i{\delta_{\pi A,0}}}{\bigl(1-\frac{2i{\Gamma^{el}}_{e}}{E-{E^{res}}_n-i\Gamma _{tot}}\bigr)},  
\label{eq: S19}
\end{equation} 
where $E^{res}=E^{thr} + E^C_{1}\approx 35.26$MeV, and $\Gamma =0.4$ MeV. 

The $2s$-energy level in the ($\pi^{-},^{12}\text{O}$) atom is separated by $0.18$ MeV, and the distance between the higher energy levels is rapidly decreasing as $\sim 1/n^3$. Therefore, one can expect that only several low lying levels will make a noticeable contribution in $\Gamma^{e}_{tot}$. In our calculations we will consider this quantity as a free parameter to be determined from the data.  

\begin{figure}[here]
\includegraphics[scale=0.3]{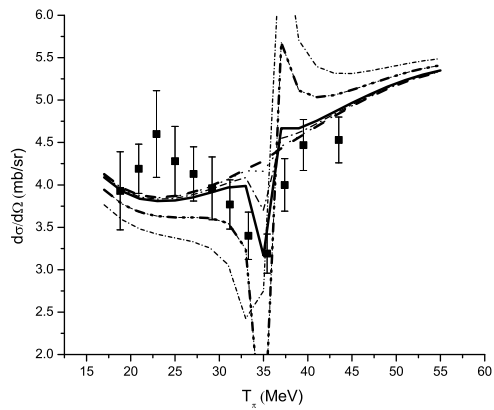}
\caption{\label{fig: fig1}The excitation function at $83^{\circ}$. Experimental data is taken from Ref. \cite{Alex98}. The dash line presents the UST calculations without taking into account the effect of sub-threshold resonances ($\gamma =0$); the dot line corresponds to $\gamma=0.01$; dash-dot  - $\gamma=0.05$ ; solid - $\gamma =0.1$; dash-dot-dot - $\gamma =0.5$; and short dash-dot - $\gamma =1.0$.}.   
\end{figure}

In Figure \ref{fig: fig1} we present calculations at $83^{\circ}$ for different values of $\gamma=\Gamma^{e}_{tot}/\Gamma$. This parameter determines the magnitude of the effective elastic width $\Gamma^{e}_{tot}$. In our calculations $\Gamma =0.4 \text{MeV}$. It is seen that the best description of the data is obtained with $\gamma \approx 0.1$ which corresponds to $\Gamma ^{e}_{tot}\approx$ 0.04 MeV.
 
\begin{figure}[h]
\includegraphics[scale=0.3]{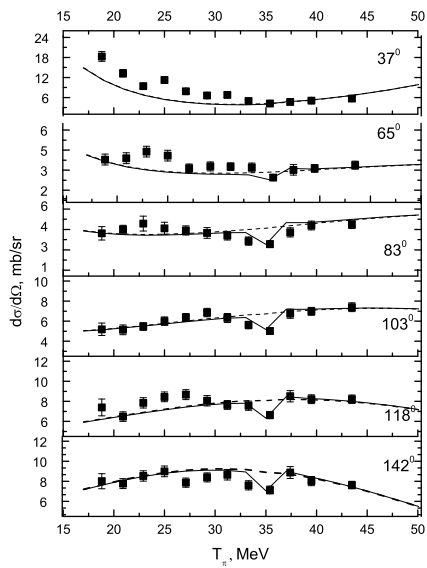}
\caption{\label{fig:fig2} Excitation functions  at fixed scattering angles. Black squares are the data from \cite{Alex98}, the lines present the UST calculations without ($\gamma=0$) and with ($\gamma=0.1$) inclusion of the sub-threshold resonance effect from formation of (${\pi}^{-}, ^{12}\text{O}$) atom.}
\end{figure}

The results of calculations of the excitation function with and without taking into account the sub-threshold resonance effect at all scattering angles measured in Ref.\cite{Alex98} are presented in Figure \ref{fig:fig2}. These calculations were performed with $\gamma=0.1$ ($\Gamma ^{e}_{tot}\approx$ 0.04 MeV) that was found to provide the best description of the data at $83^{\circ}$.          
  
\begin{figure}[h]
\includegraphics[scale=0.3]{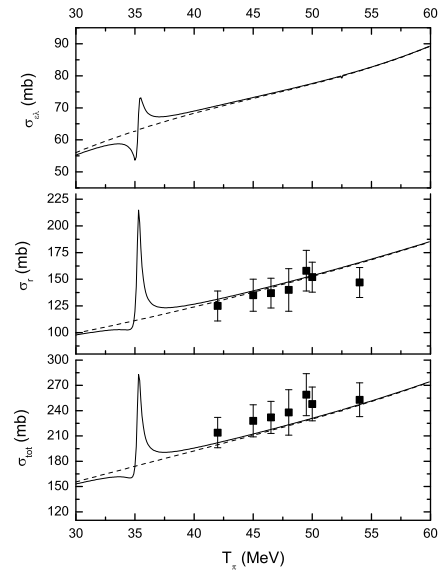}
\caption{\label{fig:fig3} Total cross sections: $\sigma _{el}, \sigma _{tot}$, and $\sigma _{r}= \sigma _{tot}-\sigma _{el}$. The lines present the results of the UST calculations without ($\gamma=0$) and with ($\gamma=0.1$) taking into account the sub-threshold resonance effect. The experimental data is taken from Ref. \cite{Saunders}}.
\end{figure}

Figure \ref{fig:fig3} shows the effect of the S-wave sub-threshold resonance on the total cross sections. One can see that the reaction cross section can reach the magnitude of $\sim 300 mb$.           

At each resonance energy the partial cross section reaches its "kinematic" maximum  $ \sigma_{l,{res}}=\frac{4\pi}{k^2}(2l+1)$. For example, for $T_{\pi}=35$ MeV for $l=0$ $\sigma_{0,{res}}=\frac{4\pi}{k^2}\approx 500 mb$. It means that even at low energies the reaction cross section can be quite big, comparable to the cross pion-nucleus sections at the $\Delta _{33}$ resonance region. There is no direct systematic experimental data on total cross sections at the sub-threshold resonance region. The data from Ref. \cite{Saunders} covers the energy region from 45 to 65 MeV and is in agreement with the UST calculations.  
 
\section{\label{sec:Sec4} Conclusion} 
In this paper we presented an explanation for the energy dependence in the excitation function in scattering of positive pions from $^{12}\text{C}$ at pion energies in the 30 - 35 MeV region which were observed in  Ref.\cite{Alex98}. It is shown that these oscillations can be explained by formation of the quasi-stationary pionic atom states in the vicinity of the threshold of the DCX reaction channel. The threshold energy is determined by the reaction Q-value and the pion's kinetic energy required overcoming the nuclear Coulomb barrier. In the considered case the threshold energy is about 35 MeV. 

The width of the resonance region is determined by the magnitude of the first Coulomb level of the pionic atom which is about 0.23 MeV for $(\pi^{-},^{12}\text{O})$ atom. The narrowness of this region may explain why other experimental groups (a detailed comparison of existing experimental data is given in Ref.\cite{Alex98}) did not see these oscillations. Fortunately, in Ref.\cite{Alex98} the data sets at different pion's energies included the pion's energy at 35.4 MeV.  

In presented analysis there is one free parameter - the elastic width of the sub-threshold resonances. The best description of the oscillations was obtained with $\Gamma^e_{tot}\approx 0.04$ MeV. It is important to note that this value refers to the integrated elastic width that accounts for the contribution of an infinite series of resonances at a given orbital momentum. In Section II it was shown that one can approximate the sum over all resonances by a single Breit-Wigner formula (\ref{eq:S18}) if the particle's decay width is much bigger than the corresponding elastic widths. In the considered reaction this condition is satisfied since the $^{12}\text{O}$ decay width is $\sim 0.4$ MeV. 

As the pion energy approaches the sub-threshold resonance energies the reaction cross section varies significantly as it is shown on Figure (\ref{fig:fig3}). It means that, despite the fact that the DCX reaction cross section is quite small by itself at low energies, the sub-threshold resonance effect amplifies the DCX role in  pion-nucleus dynamics. In addition, if the decay width of the nucleus created due to the DCX is much bigger than the corresponding elastic width the final state of the quasi-bound system is determined  by the nuclear decay. In other words, one can say that positive pions may become effective "burners" of the nuclei when their energy matches the energy of sub-threshold resonances caused by the formation of pionic atom states below the threshold of the DCX reaction.                      

\begin{acknowledgements}
The author is indebted to V. B. Belyaev and J.R. Peterson for stimulating discussions and helpful advises.     
\end{acknowledgements}

\nocite{*}

\bibliography{mkh2}


\end{document}